\documentstyle[12pt,aasms4]{article}

\received{}
\accepted{}
\journalid{}{}
\articleid{}{}

\slugcomment{To appear in ApJ Letters}

\lefthead{Zapatero Osorio et al.}
\righthead{New Brown Dwarfs in the Pleiades Cluster}

\begin{document}

\title{New Brown Dwarfs in the Pleiades Cluster}

\author{M. R. Zapatero Osorio, R. Rebolo, E. L. Mart\'\i n\altaffilmark{1}}
\affil{Instituto de Astrof\'\i{}sica de Canarias, E-38200 La Laguna,  
Tenerife. Spain}

\author{G. Basri}
\affil{Department of Astronomy, University of California, Berkeley, 
Berkeley, CA 94720. USA}

\author{A. Magazz\`u}
\affil{Osservatorio Astrofisico di Catania, Citt\`a Universitaria, I-95125 Catania, Italy}

\and

\author{S. T. Hodgkin, R. F. Jameson,  M. R. Cossburn}
\affil{Astronomy Department, Leicester University, Leicester LE1 7RH, UK}

\altaffiltext{1}{Also at: Department of Astronomy, University of California, 
Berkeley, Berkeley, CA 94720. USA}

\centerline{e-mail addresses: mosorio@iac.es, rrl@iac.es, ege@iac.es} 
\centerline{basri@soleil.berkeley.edu, antonio@ct.astro.it}
\centerline{sth@star.le.ac.uk, rfj@star.le.ac.uk, mrc@star.le.ac.uk}

\begin{abstract}
We present intermediate and low resolution optical spectroscopy 
(650--915~nm) of seven faint, very red objects ($20 > I \ge 17.8$, 
$I-Z \ge$~0.5) discovered in a CCD-based $IZ$ survey covering an area 
of 1~deg$^2$ in the central region of the Pleiades open cluster. The 
observed spectra show that these objects are very cool dwarfs, 
having spectral types in the range M6--M9. Five out of the seven 
objects can be considered as Pleiades members on the basis of their 
radial velocities, H$\alpha$ emission and other gravity sensitive 
atomic features like the Na\,{\sc i} doublet at 818.3 and 819.5~nm. 
According to current 
evolutionary models the masses of these new objects range from roughly 
80~$M_{\rm Jup}$ for the hottest in the sample down to 45~$M_{\rm Jup}$ 
for Roque~4, the coolest and faintest confirmed member. These observations 
prove that the cloud fragmentation process extends well into 
the brown dwarf realm, suggesting a rise in the initial mass function 
below the substellar limit. 

\end{abstract}


\section{Introduction}

In the last two years, with the discoveries of the first bona-fide 
brown dwarfs (BDs; Rebolo, Zapatero Osorio \& Mart\'\i n \cite{rebolo95}; 
Nakajima et al. \cite{nakajima95}), it has been proved that objects with 
masses between those of stars and planets can also be formed in nature. 
Because of its youth and proximity the Pleiades star cluster is an 
ideal hunting ground for substellar objects (see Hambly \cite{hambly97} 
for a review). The discovery of BDs like Teide~1 and Calar~3 in a small 
survey of the Pleiades  (Zapatero Osorio et al. \cite{osorio97a}) 
suggests that a large number of very low mass objects may populate this 
cluster. If this were the case, astronomers would have an unique 
opportunity to establish the observational properties of these rather 
elusive objects and to characterize the initial mass function beyond 
the star-BD boundary.

With the aim of searching for new Pleiades BDs, Zapatero Osorio et al. 
(\cite{osorio97b}) have performed  a deep CCD $IZ$ survey covering 
1~deg$^2$ within the central region of the cluster. More than 40 faint 
($I \ge 17.5$), very red ($I-Z \ge 0.5$) objects have been detected down 
to $I \sim 22$. Their location in the $IZ$  color diagram suggests cluster 
membership. According to the ``NextGen'' theoretical 
evolutionary models of Chabrier, Baraffe \& Plez (\cite{chabrier96}), 
they should have masses in the interval 80--30~$M_{\rm Jup}$ 
(1~$M_{\rm Jup} \sim 10^{-3}$~$M_{\odot}$). In this paper we present 
spectroscopic observations for seven of the candidates with magnitudes 
in the interval $I$~=~17.8--20~mag. We have determined spectral types, radial 
velocities, and H$\alpha$ emission which allow us to assess their membership, 
and therefore, their substellar nature.

\section{Observations and results}

We have collected intermediate and low resolution spectra in optical 
wavelengths for the objects listed in Table~\ref{tab1} (the full name of 
the objects is Roque Pleiades; from now on abreviated simply to Roque) 
using the William Herschel Telescope (WHT, Observatorio del Roque de los 
Muchachos, La Palma) and the KeckII telescope (Mauna Kea Observatory, 
Hawaii). Table~\ref{tab1} summarizes the log of the observations. Finding 
charts for these objects are provided in Zapatero Osorio et al. 
(\cite{osorio97b}). Figure~\ref{fig1} depicts the color-magnitude diagram 
of our 1~deg$^2$ survey in which the locations of the new BD candidates are 
indicated. Our targets were chosen to be fainter than HHJ~3 (Hambly, Hawkins 
\& Jameson \cite{hambly93}) and with ($I-Z$) colors redder than those given 
by an extrapolation of the borderline denoting the separation between 
cluster members and field objects. 

The instrumentation used was the ISIS double-arm spectrograph at the WHT 
(we only used the red arm) with the grating R158R and a TEK 
(1024$\times$1024~pix$^2$) CCD detector; and the LRIS spectrograph 
with the 830 and 1200~g~mm$^{-1}$ gratings and the TEK 
(2048$\times$2048~pix$^2$) CCD detector at the KeckII telescope. The nominal 
dispersions and the wavelength coverage provided by each instrumental 
setup are listed in Table~\ref{tab1}. Slit width projections were typically 
3~pixels, except for the observations of Roque~14 and~15, for which the 
seeing conditions forced us to have a slit width projecting onto 5 pixels. 
Exposure times ranged from 30~min to 1~hr for the faintest objects. 
Spectra were reduced by a standard procedure using IRAF\footnote{IRAF is 
distributed by National Optical Astronomy Observatories, which is operated 
by the Association of Universities for Research in Astronomy, Inc., under 
contract with the National Science Foundation.}, which included debiasing, 
flat-fielding, optimal extraction, and wavelength calibration using the 
sky lines appearing in each individual spectrum (Osterbrock et al. 
\cite{osterbrock97}). Finally, the spectra were flux-calibrated making 
use of the standards HD~19445 (WHT) and HD~84937 (KeckII), which have 
absolute flux data available in the IRAF environment. The final spectra 
are presented in Figure~\ref{fig2} together with comparison spectra of 
PPl~15 (M6.5) and Calar~3 (M8) obtained with the same instrumental 
configuration.

The observed spectra clearly correspond to very late M dwarfs 
showing prominent VO and TiO molecular absorption bands and rather 
strong atomic lines of K\,{\sc i} (766.5 and 769.9~nm) and Na\,{\sc i} 
(818.3 and 819.5~nm). In Table~\ref{tab2} we give accurate spectral types 
derived by measures of the pseudocontinuum PC1--4 indices (Mart\'\i n, 
Rebolo \& Zapatero Osorio \cite{martin96}), as well as the derived radial 
velocities and the equivalent widths of some atomic lines (H$\alpha$ and 
Na\,{\sc i}) present in the spectra. The VO index (Kirkpatrick, Henry \& 
Simons \cite{kirk95}), also measured, was found to be consistent with 
late spectral types (M6--M9). Radial velocities were obtained by 
cross-correlating the spectra in the regions 654--700~nm, 730--750~nm 
and 840--880~nm with templates observed using the same 
instrumental configuration. These templates were LHS~248 (M6.5V, Basri 
\& Marcy \cite{basri95}) and Calar~3 (M8, Rebolo et al. \cite{rebolo96}). 
We have not measured radial velocities for Roque~14 and~15 because the 
resolution of their spectra is rather low ($\sim$15~\AA). We note that 
for Roque~11 spectra of two different resolutions ($\sim$2 and 6~\AA) 
are available, and the radial velocities agree well with each other. 

There are several spectroscopic indicators that allow us to investigate 
the membership of our objects in the Pleiades. Cluster members are found 
with radial velocities in the range 0--14~km~s$^{-1}$ (Stauffer et al. 
\cite{stauffer94}). All our candidates with a radial velocity measurement 
clearly meet this criterion within the estimated error bars. Further 
evidence for membership is given by the presence of H$\alpha$ in emission. 
According to Stauffer et al. (\cite{stauffer94}) and Hodgkin, Jameson 
\& Steele (\cite{hodgkin95}), H$\alpha$ equivalent widths among very cool 
cluster members seem to be greater than 3~\AA. We find that all of our 
targets, except Roque~4, share this characteristic, which supports their 
membership. The lack of H$\alpha$ in emission in Roque~4 (M9) should not be 
interpreted as inconsistency with membership, because  the behaviour of this 
line for Pleiads later than M8 is unknown. It could be that beyond a certain 
temperature, activity decreases considerably in the atmospheres of such 
cool objects, and H$\alpha$ may be no longer seen in emission. The 
sensitivity of the Na\,{\sc i} doublet to gravity makes it useful as a 
membership criterion  (Steele \& Jameson \cite{steele95}; Mart\'\i n et 
al. \cite{martin96}). We find that the equivalent width of this doublet 
is lower in our objects than in field stars with similar spectral type 
and temperature, indicating a lower gravity and hence, a younger age. 
Finally, as expected for true Pleiads, our objects nicely fit and extend 
the sequence delineated by very low-mass cluster members in the 
$I$-magnitude versus spectral type diagram of Mart\'\i n et al. 
(\cite{martin96}). 

We conclude on the basis of all the above spectroscopic membership 
criteria that five of our objects (Roque~4, 11, 13, 16 and~17) are 
very likely members of the cluster. Additional support is provided by 
our $K$-band measurements which locate them in the Pleiades IR photometric 
sequence (Zapatero Osorio, Mart\'\i n \& Rebolo \cite{osorio97c}). We 
remark that Roque~11 is a photometric and spectroscopic ``twin'' of Teide~1 
and Calar~3, and that Roque~4 could be the least luminous and coolest cluster 
member yet found being very similar in appearance to PIZ~1 (Cossburn et al. 
\cite{cossburn97}). These two Roque objects 
together with Roque~13 can be classified as genuine BDs, while Roque~16 
and~17 might be transition objects lying in the region between stars and 
BDs. To reach a definitive conclusion on the membership status of Roque~14 
and~15 we shall await radial velocity measurements and IR photometry.

We find higher VO indices and more intense TiO molecular bands in the M8 
and M9 Pleiads than in the field dwarfs with the same spectral type. 
This might be associated with the formation of dust in the atmospheres 
(Tsuji, Ohnaka \& Aoki \cite{tsuji96}) and its dependence on gravity. 
The larger the gravity is, the larger the pressure, which favors the 
formation of grains at cool temperatures. Young BDs have lower gravities 
than field objects and therefore, dust molecules (silicates, grains) may 
condense less efficiently. The effect of grain formation is to decrease the 
number of vanadium, titanium and oxygen atoms in gas phase and therefore 
the abundance of the molecular species of VO and TiO, which results in a 
more transparent atmosphere in field dwarfs than in young BDs at optical 
wavelengths. 

\section {Discussion and final remarks}

In order to estimate the mass of our objects, we must first derive their 
luminosities. One can convert $IK$ photometry and spectral type to bolometric 
luminosity by employing relationships derived for cool field dwarfs 
(Jones et al. \cite{jones94}, and references therein), and averaging the 
results. Good agreement ($\pm$0.15~dex) has been found 
between luminosities derived from different calibrations. We have adopted 
a distance modulus of 5.53 to the Pleiades, an extinction of 
$A_{\rm I}$~=~0.07~mag, and M$_{\rm bol}$~=~4.76~mag for the Sun. 
The resulting luminosities are given in Table~\ref{tab2}. Masses have 
been inferred by comparing these luminosities 
with the theoretical evolutionary tracks for an isochrone of 120~Myr 
provided by Chabrier et al. (\cite{chabrier96}). We find that Roque~16 
and~17 have masses in the range 80--60~$M_{\rm Jup}$, similar to PPl~15 
(Basri, Marcy \& Graham \cite{basri96}), and thus, may help to define the 
star-BD boundary in the Pleiades cluster. Roque~13 has a mass inbetween 
PPl~15 and Teide~1. Roque~11 resembles Teide~1 and Calar~3, 
and hence we infer the same mass (55$\pm$15~$M_{\rm Jup}$, Rebolo et 
al. \cite{rebolo96}). Since Roque~4 is 0.2~dex less luminous than Roque~11 
its mass is 10~$M_{\rm Jup}$ smaller according to the same models, and 
thus it is the least massive BD in our sample. An object with similar 
photometric and spectroscopic characteristics, PIZ~1, has been found by   
Cossburn et al. (\cite{cossburn97}), although it still lacks a radial velocity 
measurement. Recently, the Hipparcos satellite has provided new parallax 
measurements deriving a Pleiades distance modulus of 5.32~mag 
(van Leeuwen \& Hansen-Ruiz \cite{leeuwen97}). This would impose a 
reduction in our luminosities by 0.08~dex. Lower luminosities should lead 
to an older cluster age (up to about 130--150~Myr). However, a closer 
distance and an older age roughly compensate without introducing significant 
changes in the masses determined above. 

We recall that lithium is preserved in BDs less massive than 65~$M_{\rm Jup}$ 
during their whole lifetime, in marked contrast with low mass stars 
($M \le$0.3~$M_{\odot}$) which significantly destroy this element at very 
young ages. The reappearance of lithium, although dependent on age and mass, 
should take place in quite a short luminosity range (e.g. D'Antona 
\& Mazzitelli \cite{dantona94}). At the age of the Pleiades, the lithium- 
and hydrogen-burning mass limits coincide, which makes this cluster ideal  
for characterization of the substellar borderline.  
According to theoretical predictions, Roque~11 
and~4 would have fully preserved their initial lithium content and will 
never deplete it, while our remaining higher mass BDs will be destroying or 
about to destroy their lithium. Until now only PPl~15 was considered to 
be on the borderline between BDs and stars in the Pleiades. Additional 
measurements of lithium in objects with similar characteristics are needed 
in order to provide a better location of this boundary as well as an 
improved age determination for the cluster (Basri et al. \cite{basri96}). 
The observation of lithium in Roque~4, which is cooler than Teide~1, 
is also important as it would give key information on the formation of 
lithium lines in the atmospheres of very cool dwarfs. This is a subject 
of increasing importance given the detections of lithium in the 
recently discovered extremely cool field dwarfs (Ruiz, Leggett \& Allard 
\cite{ruiz97}; Mart\'\i n et al. \cite{martin97a}; 
Tinney, Delfosse \& Forveille \cite{tinney97}; Rebolo et al. \cite{rebolo98}).

Only a small fraction ($\sim$17\%) of the BD candidates found in our $IZ$ 
survey (see Fig.~\ref{fig1}) have been investigated in this paper. We have 
collected follow-up low-resolution optical spectroscopy for seven of 
the brighter candidates ($I<20$). These observations confirm cluster 
membership for five of them, and indicate that the other two are likely 
members, although radial velocity measurementes are still required.
The number of remaining candidates in the explored area is large enough 
to ensure that follow-up spectrocopic and infrared observations will 
confirm many more BDs. Among the faintest ones, there could be BDs with 
masses as low as 30~$M_{\rm Jup}$. 

Our spectroscopic results show that a high percentage of the objects found 
in Zapatero Osorio et al.'s (\cite{osorio97b}) photometric survey in the 
Pleiades may indeed be true cluster members. The number of BD candidates 
identified indicates a continuing rise of the initial mass function 
(IMF, d$N(m)$/d$M \sim M^{-\alpha}$) across the stellar-substellar boundary. 
A preliminary estimate of the slope index can be found in Mart\'\i n et al. 
(\cite{martin97b}) which gives $\alpha$~=~1$\pm$0.5. A similar IMF slope 
was found by Meusinger, Schilbach \& Souchay (\cite{meusinger96})  
for Pleiades members with masses in the range 1.0--0.4~M$_\odot$, 
and by Hambly \& Jameson (\cite{hambly91}) for the range 0.5--0.1~M$_\odot$.  
Even though the IMF appears to rise up to about 45~$M_{\rm Jup}$, 
it is not steep enough for BDs in the mass range 
75--45~$M_{\rm Jup}$ to make a significant contribution to 
the total mass of the cluster. However, their population is probably 
quite numerous, being 200--300 in the whole cluster area. 
If the IMF is extrapolated toward very low masses, say 10~$M_{\rm Jup}$ 
(roughly the deuterium burning limit), the total number of BDs 
in the cluster would be increased to the order of 1000~objects, and 
thus BDs may even double the number of known members in the Pleiades. 
Nevertheless, they would not contribute significantly to the mass of the 
cluster (providing less than 5\% \ the mass contained in stars). Assuming 
that the substellar Pleiades IMF is representative of field objects, 
and normalizing to the local volume density of M0--M8 dwarfs identified 
within $d$~=~5~pc (0.0726 stars~pc$^{-3}$, Lang \cite{lang92}), we 
find that in the solar neighborhood BDs with masses 80--40~$M_{\rm Jup}$ 
could be as numerous as M-type dwarfs. 

\acknowledgments

{\it Acknowledgments}: This work is based on observations obtained at the 
W.~M. Keck Observatory, which is operated jointly by the University of 
California and CALTECH; and at the WHT telescope operated by the Isaac 
Newton Group of Telescopes funded by PPARC at the ORM. This work has been 
supported by the European Commission through the Activity ``Access to 
Large-Scale Facilities'' within the Programme TMR, awarded to the 
Instituto de Astrof\'\i sica de Canarias to fund European Astronomers 
access to the Canary Islands Observatories (European Northern Observatory). 
Partial support has been provided by the Spanish DGES project no. 
PB95-1132-C02-01.

\clearpage

\figcaption[f1.eps]{\label{fig1} The $I$ vs ($I-Z$) diagram resulting 
from our survey covering 1~deg$^2$ in the central region of the Pleiades 
cluster. Filled symbols represent the BD candidates 
whose spectra we present in this paper. Masses (right side, $M_{\rm Jup}$ 
units) correspond to 120~Myr (Chabrier et al. 1996) at the Pleiades 
distance of 127~pc.}

\figcaption[f2.ps]{\label{fig2} Low resolution spectra (4--15~\AA) 
obtained for our Pleiades BD candidates and for the BDs PPl~15 and 
Calar~3 (top panel: WHT spectra; low panel: KeckII spectra) . 
Spectral types range from M6 to M9. Some atomic and molecular 
features are indicated. The normalization point has been taken at 813~nm. 
A constant step of 0.5 units has been added to each spectrum for clarity.}

\clearpage

\begin{deluxetable}{lcccc}
\tablecaption{\label{tab1} Log of spectroscopic observations}
\tablewidth{0pt}
\tablehead{
\colhead{Objects} & \colhead{Teles.} & \colhead{Disp.}           
& \colhead{$\Delta \lambda$} & \colhead{Date}  \nl
\colhead{}        & \colhead{}       & \colhead{(\AA~pix$^{-1}$)}
& \colhead{(nm)}             & \colhead{(1996)} }
\startdata
Roque~4 \& 11; Calar~3             & KeckII & 1.83 & 654--833 & Dec 3    \nl
Roque~11                           & KeckII & 0.63 & 654--775 & Dec 4    \nl
Roque~13, 14, 15, 16 \& 17; PPl~15 & WHT    & 2.90 & 650--915 & Dec 8--9 \nl
\enddata
\end{deluxetable}

\begin{deluxetable}{lcccclcrcc}
\footnotesize
\tablecaption{\label{tab2} Data for our Pleiades BDs}
\tablewidth{0pt}
\tablehead{
\colhead{Name}  & 
\colhead{RA$_{\rm J2000}$}    & 
\colhead{DEC$_{\rm J2000}$}   & 
\colhead{$I$}   & 
\colhead{$I-K$} & 
\colhead{SpT}   & 
\colhead{NaI}   & 
\colhead{H$\alpha$} & 
\colhead{$v_{\rm rad}$} & 
\colhead{log~$L/L_{\odot}$}   \nl
\colhead{}      & 
\colhead{($^{\rm h \ m \ s}$)} & 
\colhead{($^{\circ} \ ' \ ''$)} & 
\colhead{}      & 
\colhead{}      & 
\colhead{}      & 
\colhead{(\AA)} & 
\colhead{(\AA)} &        
\colhead{(km s$^{-1}$)} &
                            }
\startdata
Roque~16 & 3 47 39.0 & 24 36 22 & 17.79 & 3.18 & M6  & 4.7 & 5.0 & 
$-20\pm15$ & $-2.89$  \nl
Roque~15 & 3 45 41.2 & 23 54 11 & 17.82 &      & M6.5& 6.0 & 4.0 & 
           & $-2.86$  \nl
Roque~17 & 3 47 23.9 & 22 42 38 & 17.78 & 3.45 & M6.5& 4.5 &15.0 & 
$-14\pm15$ & $-2.83$  \nl
Roque~14 & 3 46 42.9 & 24 24 50 & 18.21 &      & M7  & 5.0 &17.0 & 
           & $-3.00$  \nl
Roque~13 & 3 45 50.6 & 24 09 03 & 18.25 & 3.65 & M7.5& 5.4 &10.5 & 
$-1\pm15$  & $-3.00$  \nl
Roque~11 & 3 47 12.1 & 24 28 32 & 18.73 & 3.63 & M8  & 4.8 & 5.8 & 
$-6\pm12$  & $-3.15$  \nl
         &           &          &       &      &     &     &     & 
$-3.5\pm7$\tablenotemark{a} & \nl
Roque~4  & 3 43 53.5 & 24 31 11 & 19.75 & 4.52 & M9  & 4.7 &$<$5 & 
$+4\pm12$  & $-3.35$  \nl
\enddata
\tablenotetext{a}{Radial velocity measured from the high resolution 
spectrum obtained at the KeckII telescope.}
\tablenotetext{}{Coordinates are accurate to $\pm3''$. The uncertainty 
in the spectral type determination is $\pm$0.5 subclasses. Typical error bars 
for the equivalent widths of the atomic lines is $\pm1$~\AA. Luminosities 
are given to a relative accuracy of $\pm$0.02~dex due to errors in the 
photometry. All error bars are given at 1-$\sigma$.}
\end{deluxetable}


\plotone{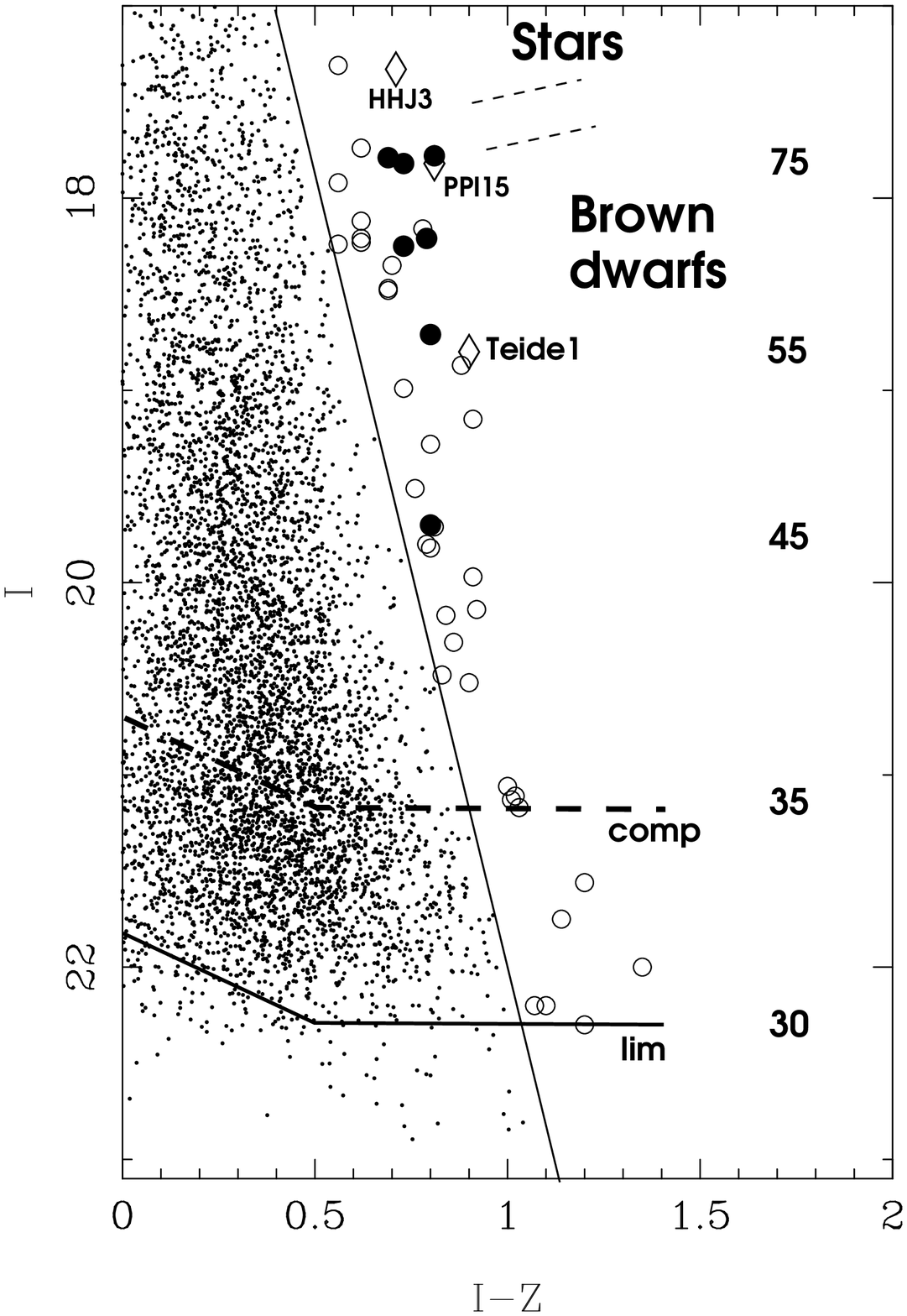}

\plotone{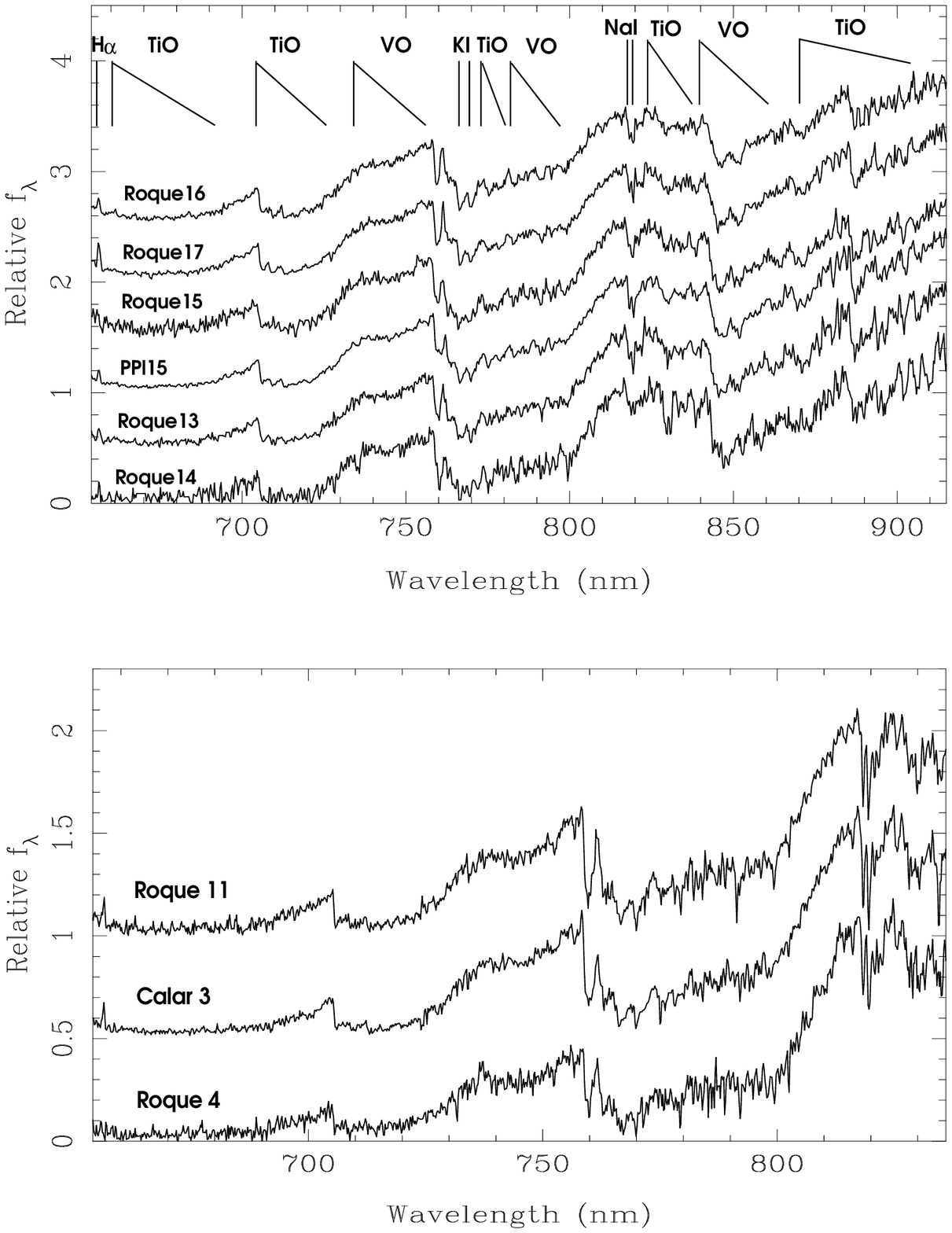}

\end{document}